\documentclass[journal, twoside, final]{IEEEtran}
\usepackage{cite}
\usepackage{graphicx}
\usepackage{amssymb, amsmath, amsthm, amsfonts}
\usepackage{caption}     %this package and the next one are used to support subfigures.
\usepackage{subcaption}
\usepackage{multirow, multicol}
\usepackage{mathrsfs}
\usepackage{color}
\usepackage{url}
\usepackage{flushend}
\usepackage{setspace}
\usepackage{placeins}
\usepackage{booktabs}
\allowdisplaybreaks

\usepackage{epstopdf}

\usepackage{longtable}

\usepackage{pdflscape}

\allowdisplaybreaks
\usepackage{array}

\newtheorem{lemma}{Lemma}
\newtheorem{theorem}{Theorem}

\newtheorem{corollary}{Corollary}

\begin{document}

\title{\Huge Unified Analytical Volume Distribution of Poisson-Delaunay Simplex and its Application to Coordinated Multi-Point Transmission}
\author{Minghua~Xia,~\IEEEmembership{Member,~IEEE}, and Sonia~A\"{i}ssa,~\IEEEmembership{Senior Member,~IEEE}
\thanks{Manuscript received October 23, 2017; revised March 12, 2018 and April 29, 2018; accepted May 1, 2018. This work was supported by the National Natural Science Foundation of China under Grant 61671488. The associate editor coordinating the review of this paper and approving it for publication was X. Zhou.}
\thanks{M. Xia is with the School of Electronics and Information Technology, Sun Yat-sen University, Guangzhou, 510006, China  (e-mail: xiamingh@mail.sysu.edu.cn).}
\thanks{S. A\"{i}ssa is with the Institut National de la Recherche Scientifique (INRS), University of Quebec, Montreal, QC, H5A 1K6, Canada (e-mail: aissa@emt.inrs.ca).}
\thanks{Color versions of one or more of the figures in this paper are available online at http://ieeexplore.ieee.org.}
\thanks{Digital Object Identifier }
}

\markboth{IEEE Transactions on Wireless Communications} {Xia \MakeLowercase{\textit{et al.}}: Unified Analytical Volume Distribution of Poisson-Delaunay Simplex and its Application to Coordinated Multi-Point Transmission}

\maketitle

\IEEEpubidadjcol

\IEEEpubid{\begin{minipage}{\textwidth} \ \\[12pt] \centering 1536-1276 \copyright\ 2018 IEEE. Translations and content mining are permitted for academic research only. Personal use is also permitted, \\
but republication/redistribution requires IEEE permission. See \url{http://www.ieee.org/publications_standards/publications/rights/index.html for more information}.\end{minipage}}

\begin{abstract}
    \noindent For Poisson-Delaunay triangulations in $d$-dimensional Euclidean space $\mathbb{R}^{d}$, a structured and computationally efficient form of the probability density function (PDF) of the volume of a typical cell is analytically derived in this paper. In particular, the ensuing PDF and the corresponding cumulative density function (CDF) are exact and unified, applicable to spaces of arbitrary dimension ($d \ge 1$). Then, the special cases and shape characteristics of the resulting PDF are thoroughly examined. Finally, various applications of the obtained distribution functions are outlined and, in particular, a novel coordinated multi-point transmission scheme based on Poisson-Delaunay triangulation is developed and the pertinent void cell effect is precisely evaluated by using the obtained distribution functions.
\end{abstract}

\begin{IEEEkeywords}
\noindent Meijer's $\mathrm{G}$-function, Poisson-Delaunay triangulation, Poisson-Voronoi tessellation, stochastic geometry, void cell effect, volume distribution.
\end{IEEEkeywords}

\section{Introduction}
\label{Section:Introduction}
Voronoi tessellations and their corresponding cellular structures in $d$-dimensional Euclidean space $\mathbb{R}^{d}$, for all $d \ge 1$, are encountered in many scientific fields, e.g., astronomy, biology, crystallography and ecology \cite{Okabe00}. The oldest documented trace of their application dates back to Johannes Kepler (1571--1630) in his study on the shapes of snowflakes and the densest sphere packing problem \cite{Liebling12}. In more recent decades, Voronoi tessellations have become cornerstones of modern disciplines such as computational geometry, algorithm design, scientific computing, optimization,  information theory and wireless communications \cite{Chiu13}. For instance, the operation of the well-known maximum likelihood decoding is essentially equivalent to defining the Voronoi tessellation of the set of transmitted codewords and finding the Voronoi cell in which each received codeword is located. Another similar example pertains to vector quantization.

\IEEEpubidadjcol

Among various Voronoi tessellation types, the one based on homogeneous Poisson point processes (PPPs) is the most basic and useful. Such tessellations are termed Poisson-Voronoi tessellations and their dual graphs are the so-called Poisson-Delaunay triangulations. Compared with Poisson-Voronoi tessellations, the probability distributions of the characteristics of a typical cell of Poisson-Delaunay triangulations (or equivalently a Poisson-Delaunay simplex) are mathematically more tractable and, thus, the properties and statistics of a typical cell of the latter triangulations were extensively studied and are well documented in the encyclopedic monograph \cite[Section 5.11]{Okabe00}. However, as one of the fundamental features, knowledge of the exact probability distributions of the volume of a typical Poisson-Delaunay cell in $\mathbb{R}^{d}$ is very limited so far. Specifically, in the seminal work by P. N. Rathie \cite{RathieJAP92} in 1992, although a general Mellin-Barne integral representation for the probability density function (PDF) of the volume of a typical cell was developed, it is too complex to be further processed. Actually, only if in the particular cases with $d=1$ and $d=2$, analytical expressions for the PDF of the volume of a typical cell were obtained individually. For the special case with $d=3$, however, Rathie's expression is extremely complicated, due to three infinite series involved, consisting of Gamma, Psi and Zeta functions. In the same context, the other notable work is by L. Muche in 1996, where a three-fold integral expression for the volume's PDF of a typical cell in the particular case of $d=3$ was reported \cite{MucheJSP96}. For higher dimensional cases with $d \ge 4$, no any analytical expression for the distribution function of the volume was ever reported in the open literature, which hinders further theoretical development and practical applications of the powerful Poisson-Delaunay triangulations. For the interested reader, the state-of-the-art on the volume distribution of a typical cell of Poisson-Delaunay triangulations can be found in the latest version of the classic reference \cite[Section 9.7.4]{Chiu13}.

\IEEEpubidadjcol

In this contribution, analytical expressions for the PDF and cumulative distribution function (CDF) of the volume of a typical cell of Poisson-Delaunay triangulations are originally developed. The resulting expressions are unified and can be applied in arbitrary $d$-dimensional Euclidean space $\mathbb{R}^{d}$ with $d \ge 1$. In particular, the obtained distribution functions include the aforementioned ones reported in the open literature as special cases. Furthermore, since the major results are expressed in terms of the Meijer's $\mathrm{G}$-function, they can be readily computed in a numerical way by using built-in functions in popular numerical softwares, such as the function $\mathsf{MeijerG}$ in Mathematica$^\circledR$ or $\mathsf{meijerG}$ in Matlab$^\circledR$. As an illustrating application of the obtained mathematical results to wireless communications, a novel coordinated multi-point (CoMP) transmission scheme based on Poisson-Delaunay triangulation is designed, and its pertinent void cell effect is precisely evaluated by using the obtained distribution functions.

The rest of this paper is organized as follows. Section \ref{Section-Moments} starts with a rigorous definition of Poisson-Delaunay triangulations and the moments of the volume of a typical cell. The unified analytical volume distributions of a typical cell are then developed in Section~\ref{Section:MainResults}. In Section~\ref{Section:SpecialCases}, special cases of the resulting unified distribution functions are discussed and compared with previously reported results in the literature. Also, the shape characteristics of the resulting PDF are examined. Afterwards, in Section~\ref{Apps} various applications of the obtained results are outlined and, in particular, an application to CoMP transmission in wireless communications is elaborated. Finally, concluding remarks are provided in Section~\ref{Section:Conclusions}.

\section{Moments of the Typical Poisson-Delaunay Cell}
\label{Section-Moments}
For completeness of mathematical exposition, we start with the definition of Poisson-Delaunay triangulations. Let $X$ be a stationary Poisson point process with intensity $\rho$, in $\mathbb{R}^{d}$. Any $d +1$ points $\{x_{0}, \cdots, x_{d}\}$ of $X$ define almost surely a unique open ball containing these points on its boundary. If the interior of the ball contains no other point of $X$, the simplex $\mathrm{conv}\{x_{0}, \cdots, x_{d}\}$ is called a Poisson-Delaunay cell, where the operator $\mathrm{conv}\{\cdot\}$ means convex hull. The collection of all cells obtained in this way, denoted $Y$, is called the Poisson-Delaunay tessellation induced by $X$. Poisson-Delaunay tessellation is also widely known as Poisson-Delaunay triangulation, since its component cell is of a triangular form in the planar case, i.e., when $d = 2$.

According to \cite[Corollary 7.6]{MollerAAP8901}, the $k^{\mathrm{th}}$-order moment of the volume $V$ of a typical Poisson-Delaunay cell in arbitrary $d$-dimensional Euclidean space $\mathbb{R}^{d}$, can be written as Eq.~\eqref{Eq-moments} on the top of the next page, where the operator $\mathcal{E}\{\cdot\} $ means mathematical expectation and $\Gamma(x) = \int_{0}^{\infty}{t^{x-1}\exp(-t)}\,\mathrm{d}t$ denotes the Gamma function. Based on the moments shown in \eqref{Eq-moments}, exact distribution functions of $V$ (PDF and CDF), applicable to any case with $d \ge 1$, are explicitly derived in the sequel, by using the methodology of inverse Mellin transform. Before delving into the details of the derivations, however, the uniqueness of the CDF of $V$ has to be guaranteed since a given set of moments does not necessarily determine a unique distribution function. For instance, the lognormal distribution is not determined by its moments \cite{HeydeJRSS6302}.

It is noteworthy that the so-called `typical cell' is not a particular cell chosen from a given tessellation. In fact, it can be seen as a generic cell sampled from the population of all cells by random selection, whereby all cells have the same chance to be chosen. For the interested reader, a rigorous mathematical definition of a typical cell of random tessellations can be found in \cite[p.~450]{Schneider08}.

\newcounter{mytempeqncnt1}
\begin{figure*}[htp]
\setcounter{mytempeqncnt1}{\value{equation}}
\setcounter{equation}{0}
\begin{equation}  \label{Eq-moments}
\mathcal{E}\{V_d^k\}
= \frac{\Gamma(d+k) \, \Gamma\left(\frac{d^2}{2}\right)\Gamma\left(\frac{d^2+dk+k+1}{2}\right)\Gamma^{d-k+1}\left(\frac{d+1}{2}\right)}
{(2^{d}\,\pi^{(d-1)/2}\rho)^k \, \Gamma(d) \, \Gamma\left(\frac{d^2+1}{2}\right)\Gamma\left(\frac{d^2+dk}{2}\right)\Gamma^{d+1}\left(\frac{d+k+1}{2}\right)}
\,\prod_{i=2}^{d+1}{\frac{\Gamma\left(\frac{k+i}{2}\right)}{\Gamma\left(\frac{i}{2}\right)}} , \, \quad k=1, 2, 3, \cdots
\end{equation}
\end{figure*}
\setcounter{equation}{1}

\section{Unified Distribution Functions of the Volume}
\label{Section:MainResults}
In the following lemma, the uniqueness of the CDF of the volume with the given set of moments shown in \eqref{Eq-moments} on the top of the next page is guaranteed.
\begin{lemma} \label{lemma-1}
The given set of moments shown in \eqref{Eq-moments} determines a CDF uniquely.
\end{lemma}
\begin{IEEEproof}
The main ingredient of the proof is \cite[Theorem 3.6.1]{Springer79}, which states that the set of moments shown in \eqref{Eq-moments} determines a CDF uniquely if the series $\sum_{k=0}^{\infty}\mathcal{E}\{V_d^k\} \frac{t^{k}}{k!}$ converges for some real non-zero variable $t$. Hence, the remaining task is to demonstrate the convergence of the aforementioned series by using the standard ratio test, which is detailed in \cite{RathieJAP92}.
\end{IEEEproof}

\newcounter{mytempeqncnt2}
\begin{figure*}[!t]
\setcounter{mytempeqncnt2}{\value{equation}}
\setcounter{equation}{1}
\begin{equation}  \label{Eq-PDF}
f_V(x)
= \frac{A}{x} \, \mathrm{G}_{p,\, q}^{m,\, 0}\left[ B x^2 \left \vert \begin{gathered}
\underbrace{\frac{d}{2}+\frac{1}{d}, \, \frac{d}{2}+\frac{2}{d}, \, \cdots, \, \frac{d}{2}+\frac{d-1}{d}}_{(d-1) \, {\rm terms}}, \,
\underbrace{\frac{d+1}{2}, \, \frac{d+1}{2}, \, \cdots, \, \frac{d+1}{2}}_{(d-1) \, {\rm terms}}
\\
\underbrace{\frac{2}{2}, \, \frac{3}{2}, \, \cdots, \, \frac{d}{2}}_{(d-1) \, {\rm terms}}, \,
\underbrace{\frac{d^2+1}{2(d+1)}, \, \frac{d^2+3}{2(d+1)}, \, \cdots, \, \frac{d^2+2d+1}{2(d+1)}}_{(d+1) \, {\rm terms}}
 \end{gathered}\right.\right].
\end{equation}
\newline
\begin{equation}  \label{Eq-CDF}
F_V(x)
= \frac{A}{2} \, \mathrm{G}_{p+1,\, q+1}^{m,\, 1}\left[ B x^2 \left \vert \begin{gathered}
1, \, \underbrace{\frac{d}{2}+\frac{1}{d}, \, \frac{d}{2}+\frac{2}{d}, \, \cdots, \, \frac{d}{2}+\frac{d-1}{d}}_{(d-1) \, {\rm terms}}, \,
\underbrace{\frac{d+1}{2}, \, \frac{d+1}{2}, \, \cdots, \, \frac{d+1}{2}}_{(d-1) \, {\rm terms}}
\\
\underbrace{\frac{2}{2}, \, \frac{3}{2}, \, \cdots, \, \frac{d}{2}}_{(d-1) \, {\rm terms}}, \,
\underbrace{\frac{d^2+1}{2(d+1)}, \, \frac{d^2+3}{2(d+1)}, \, \cdots, \, \frac{d^2+2d+1}{2(d+1)}}_{(d+1) \, {\rm terms}}, \, 0
 \end{gathered}\right.\right].
\end{equation}
\setcounter{equation}{\value{mytempeqncnt2}}
\end{figure*}
\setcounter{equation}{3}

In light of Lemma~\ref{lemma-1}, the main result of this paper is formulated in the following theorem.
\begin{theorem} \label{Theorem1}
The PDF and CDF of the volume of a typical cell of Poisson-Delaunay triangulations in arbitrary $d$-dimensional Euclidean space $\mathbb{R}^{d}$, for all $d \ge 1$, can be analytically given by Eqs.~\eqref{Eq-PDF} and \eqref{Eq-CDF} on the top of the next page, respectively, where $\mathrm{G}[\,.\,|\,.\,]$ denotes the Meijer's $\mathrm{G}$-function \cite[Section~16.17]{Olver10}, and
\begin{align}
m     & = 2d,       \label{Eq-CDF-a} \\
p      & = 2d-2,     \label{Eq-CDF-b} \\
q      & = 2d,        \label{Eq-CDF-c} \\
A     & = \frac{2^{d-1/2} (d+1)^{d^{2}/2} \, \Gamma\left(\frac{d^2}{2}\right)\Gamma^{d}\left(\frac{d+1}{2}\right)}
                {\pi \, d^{(d^2-1)/2} \, \Gamma(d)\, \Gamma\left(\frac{d^2+1}{2}\right) \prod\limits_{i=2}^{d}{\Gamma\left(\frac{i}{2}\right)}},  \label{Eq-CDF-d} \\
B    & = \left[\frac{2^{d-1} \pi^{(d-1)/2} d^{d/2} \, \Gamma\left(\frac{d+1}{2}\right)}{(d+1)^{(d+1)/2}}\right]^{2}.  \label{Eq-CDF-e}
\end{align}
\end{theorem}
\begin{IEEEproof}
In view of Lemma~\ref{lemma-1} and by recalling the uniqueness of the inverse Mellin transform, the PDF of the volume can be computed by
\begin{equation} \label{Eq.Proof-1}
f_{V}(x)
= \frac{1}{j2{\pi}x} \int_{L}{x^{-s} \, \mathcal{E}\{V_d^s\}} \,\mathrm{d}s,
\end{equation}
where $j=\sqrt{-1}$ denotes the imaginary unit, $L$ is a suitable Meillin-Barnes contour for which $f_{V}(x)$ can be computed, and $\mathcal{E}\{V_d^s\}$ is similarly defined by \eqref{Eq-moments} for complex values of $s$.\footnote{As far as the Meillin-Barnes integral given by Eq.~\eqref{Eq.Proof-1} is concerned, $s$ is a complex variable and the path of integration ($L$) is a straight line parallel to the imaginary axis with indentations, if necessary, to avoid poles of the integrand. In such a case, the variable $k$ taking integer values in Eq.~\eqref{Eq-moments} must be extended to take complex values and, accordingly, $k$ is replaced by~$s$. Moreover, as per the path of integration, the value of the imaginary part of~$s$ must approach infinity. By recalling the asymptotic expansion of the Gamma function, i.e., $\lim_{|y| \to \infty} |\Gamma(x+jy)| \exp\left(\frac{1}{2}\pi|y|\right)y^{\frac{1}{2}-x} = \sqrt{2\pi}$ \cite[Eq. 1.18.(6)]{Erdelyi53}, where $x$ and $y$ take real values, Eq.~\eqref{Eq.Proof-1} holds and its convergence can be further investigated. For more details, the interested reader is referred to \cite[Section 1.19]{Erdelyi53}.} Then, substituting \eqref{Eq-moments} into \eqref{Eq.Proof-1} and performing some algebraic manipulations by use of the Legendre duplication formula for the Gamma function, namely, $2^{2x-1}\Gamma(x)\,\Gamma(x+1/2)=\sqrt{\pi}\,\Gamma(2x)$ \cite[Eq.~(5.5.5)]{Olver10}, yields

\vspace{-10pt}
{\footnotesize
\begin{equation}  \label{Eq.Proof-2}
f_{V}(x)
= \frac{A}{j2{\pi}x} \int_{L}
{ \frac{\prod\limits_{i=2}^{d}\Gamma\left(\frac{i}{2}+s\right) \prod\limits_{i=0}^{d}\Gamma\left(\frac{d^2+1+2i}{2(d+1)}+s\right)}
{\prod\limits_{i=1}^{d-1}\Gamma\left(\frac{d}{2}+\frac{i}{d}+s\right) \Gamma^{d-1}\left(\frac{d+1}{2}+s\right)}
\left(B x^2\right)^{-s} } \,\mathrm{d}s,
\end{equation}
}
\hspace{-5pt}where constants $A$ and $B$ are defined by \eqref{Eq-CDF-d} and \eqref{Eq-CDF-e}, respectively.\footnote{It is noteworthy that the general integral expression given by Eq.~\eqref{Eq.Proof-2} was originally shown in Eq. (2.5) of \cite{RathieJAP92}. In \cite{RathieJAP92}, the said expression was not further processed due to its extremely high complexity, but only three particular cases were discussed separately (cf. details in Section~\ref{Section:Introduction}).} After careful observation, we recognize that \eqref{Eq.Proof-2} is indeed an integral of a Mellin-Barnes type, involving the product and ratio of Gamma functions. This type of integral can be expressed in terms of Meijer's $\mathrm{G}$-functions. Then, by comparing \eqref{Eq.Proof-2} with the definition of the Meijer's $\mathrm{G}$-function given by \cite[Definition 2.1]{Mathai93} or \cite[Section~16.17]{Olver10}, i.e.,
\begin{eqnarray}
\label{Eq-MeijerG}
	\lefteqn{\mathrm{G}_{p,\, q}^{m,\, n}\left[ x \left \vert \begin{gathered}   a_1, \cdots, a_p \\ b_1, \cdots, b_q \end{gathered}\right.\right]}  \nonumber \\
		\hspace{-8pt} & = & \hspace{-8pt} \frac{1}{j2{\pi}} \hspace{-3pt} \int_{L}\hspace{-6pt}
				{\frac{\prod\limits_{i=1}^{m}\Gamma\left(b_i+s\right) \prod\limits_{i=1}^{n}\Gamma\left(1-a_i-s\right)}
				{\prod\limits_{i=m+1}^{q}\hspace{-5pt}\Gamma\left(1-b_i-s\right) \prod\limits_{i=n+1}^{p}\hspace{-5pt}\Gamma\left(a_i+s\right)} x^{-s} }\mathrm{d}s,
\end{eqnarray}
yields the intended \eqref{Eq-PDF}.

Next, to derive \eqref{Eq-CDF}, we reformulate the PDF shown in \eqref{Eq-PDF} as
\begin{eqnarray}
f_V(x)
& = &  \frac{ABx}{B x^2} \, \mathrm{G}_{p,\, q}^{m,\, 0}\left[ B x^2 \left \vert \begin{gathered}   (a_p) \\ (b_q) \end{gathered}\right.\right]  \label{Eq-Proof-3} \\
& = &  ABx \, \mathrm{G}_{p,\, q}^{m,\, 0}\left[ B x^2 \left \vert \begin{gathered}  (a_{p}-1) \\ (b_{q}-1) \end{gathered}\right.\right], \label{Eq-Proof-4}
\end{eqnarray}
where sets $(a_p)$ and $(b_p)$ are used for short to denote the corresponding parameters shown in \eqref{Eq-PDF}, $(a_{p}-1)$ and $(b_{p}-1)$ mean that each element of the corresponding set is subtracted by unity, and where the scaling property of the Meijer's $\mathrm{G}$-function \cite[Eq.~(2.2.1)]{Mathai93} was exploited to obtain \eqref{Eq-Proof-4}.

Then, by definition and in view of \eqref{Eq-Proof-4}, the CDF of the volume can be computed as
\begin{eqnarray}  \label{Eq-Proof-3}
F_V(x)
& = &  AB\int_{0}^{x}{t \, \mathrm{G}_{p,\, q}^{m,\, 0}\left[ B t^2 \left \vert \begin{gathered}   (a_{p}-1) \\ (b_{q}-1) \end{gathered}\right.\right]} \, \mathrm{d}t  \label{Eq-Proof-5} \\
& = &  \frac{A}{2} \int_{0}^{B x^2}{ \mathrm{G}_{p,\, q}^{m,\, 0}\left[ y \left \vert \begin{gathered}   (a_{p}-1) \\ (b_{q}-1) \end{gathered}\right.\right]} \, \mathrm{d}y  \label{Eq-Proof-6} \\
& = &  \frac{A}{2} \, \mathrm{G}_{p+1,\, q+1}^{m,\, 1}\left[ B x^2 \left \vert \begin{gathered} 1, \,  (a_{p}) \\ (b_{q}), \, 0 \end{gathered}\right.\right],  \label{Eq-Proof-7}
\end{eqnarray}
where the change of variable $y=Bt^2$ was used to reach \eqref{Eq-Proof-6} from \eqref{Eq-Proof-5}, and \cite[Eq.~(14)]{XiaTCOM1211} was exploited to derive \eqref{Eq-Proof-7}, which is finally given in an explicit way by the desired \eqref{Eq-CDF}.
\end{IEEEproof}

It is observed from \eqref{Eq-PDF} that $x=0$ is a singular point of the PDF expression. In order to determine the value of the PDF when $x \to 0$, we have the following corollary.
\begin{corollary} \label{Corollary_Limit}
As $x \to 0$, the limit of the PDF of the volume of a typical cell of Poisson-Delaunay triangulations in $\mathbb{R}^{d}$, shown in \eqref{Eq-PDF}, converges to a constant if $d=1$, and to zero if $d \ge 2$, i.e.,
\begin{equation} \label{Eq.Singularity}
\lim_{x \to 0}{f_V(x)} = \left\{ \begin{array}{rl}
c, & {\rm if~}  d =1;     \\
0, & {\rm if~}  d \ge 2.
\end{array}\right.
\end{equation}
\end{corollary}

\begin{IEEEproof}
For better clarity, the asymptotical equivalence of the Meijier's $\mathrm{G}$-function related to our application is restated. Specifically, as $x \to 0$, we have \cite[p.~146]{Mathai93}
\begin{equation} \label{Eq-Equivalence}
\mathrm{G}_{p,\, q}^{m,\, 0}\left[ x \left \vert \begin{gathered}  a_1, \, \cdots, \, a_p  \\  b_1, \, \cdots, \, b_p \end{gathered}\right.\right] \sim |x|^{\alpha}, \quad p \le q,
\end{equation}
where the Landau notation $f(x) \sim g(x)$ is defined as $\lim_{x \to 0}f(x)/g(x) = c < \infty$, and $\alpha = \min\{b_j\}$ for $j=1, \cdots, m$. Then, by comparing \eqref{Eq-Equivalence} with \eqref{Eq-PDF}, we have
\begin{equation} \label{Eq-Factor}
\alpha = \left\{ \begin{array}{rl}
\frac{1}{2}, & \mbox{if~} d =1;     \\
\frac{5}{6}, & \mbox{if~} d =2;     \\
        1,        & \mbox{if~} d \ge 3. \\
\end{array}\right.
\end{equation}
Then, applying \eqref{Eq-Equivalence} and \eqref{Eq-Factor} to \eqref{Eq-PDF} yields the desired \eqref{Eq.Singularity}.
\end{IEEEproof}

Later in Section~\ref{SpecialCases}, the value of the constant $c$ involved in Corollary~\ref{Corollary_Limit} will be shown to be unity.

Further, we note that Theorem~\ref{Theorem1} is dedicated to Poisson-Delaunay triangulations with normalized intensity. For the general case with intensity $\rho \ne 1$, we have the following corollary.
\begin{corollary}   \label{Corollary_Intensity}
For Poisson-Delaunay triangulations with intensity $\rho \ne 1$, the PDF and CDF of the volume of a typical cell are given by
\begin{equation}
f_V^{\rho}(x)
=  \rho f_V({\rho}x)   \label{Eq-PDF2}
\end{equation}
and
\begin{equation}
F_V^{\rho}(x)
=  F_V({\rho}x),       \label{Eq-CDF2}
\end{equation}
where $f_V(x)$ and $F_V(x)$ are shown in \eqref{Eq-PDF} and \eqref{Eq-CDF}, respectively.
\end{corollary}
\begin{IEEEproof}
Eqs. \eqref{Eq-PDF2} and \eqref{Eq-CDF2} can be readily obtained by recalling the Mapping theorem \cite[p.~18]{Kingman93}.
\end{IEEEproof}

\section{Special Cases and Discussions}
\label{Section:SpecialCases}
In order to illustrate the effectiveness of the main results summarized in Theorem~\ref{Theorem1}, in this section the general expressions shown in \eqref{Eq-PDF} and \eqref{Eq-CDF} are applied in the particular cases of $d = 1, 2, 3$ and $d \ge 4$, and further compared with results reported in the open literature. Afterwards, general shape characteristics of the PDF are examined. To facilitate subsequent computations, the values of $A$ and $B$ involved in Theorem~\ref{Theorem1} are off-line computed and summarized in Table~\ref {Table_ParametersAB}.

\begin{table*}[!h]
\caption{Values of parameters $A$ and $B$ in Theorem~\ref{Theorem1}, computed as per \eqref{Eq-CDF-d} and \eqref{Eq-CDF-e}, respectively.}
\begin{center}
\begin{tabular}{cccccc}
\toprule
$d$         & 1                              & 2                                & 3                                        & 4                                                 & 5     \\\midrule
\midrule
$A$        & $\frac{2}{\sqrt{\pi}}$  &$\frac{3}{\sqrt{\pi}}$   &$\frac{560\sqrt{2}}{81\pi}$   &$\frac{234375}{18304\sqrt{2}}$   &$\frac{39919426911\sqrt{3}}{244140625\pi^{3/2}}$  \\
\midrule
$B$        & $\frac{1}{4}$             &$\frac{4\pi^2}{27}$     &$\frac{27\pi^2}{16}$             &$\frac{9216\pi^4}{3125}$             &$\frac{50000\pi^4}{729}$      \\
\bottomrule
\end{tabular}
\vspace{-15pt}
\end{center}
\label{Table_ParametersAB}
\end{table*}

\subsection{Special Cases}
\label{SpecialCases}
\begin{enumerate}
\item{\textbf{Case I ($\boldsymbol{d=1}$)}:} In this condition, according to Table~\ref{Table_ParametersAB}, we have $A=2/\sqrt{\pi}$ and $B=1/4$. Then, the general PDF given by \eqref{Eq-PDF} reduces to
\begin{eqnarray}
f_V(x)
&  =  &  \frac{2}{x\sqrt{\pi}} \, \mathrm{G}_{0,\, 2}^{2,\, 0}\left[ \frac{x^2}{4} \left \vert \begin{gathered}  \line(1, 0){15} \\  \frac{1}{2}, \, 1 \end{gathered}\right.\right]  \label{Eq.CaseI-1} \\
&  =  &  \frac{4}{x\sqrt{\pi}} \, \left(\frac{x^2}{4}\right)^{3/4}\, K_{-\frac{1}{2}}(x)  \label{Eq.CaseI-2} \\
&  =  &  \exp(-x),  \label{Eq.CaseI-3}
\end{eqnarray}
where $\mathrm{G}_{0,\, 2}^{2,\, 0}\left[x \left \vert  ^{\line(1, 0){15}} _{b_1 , \, b_2} \right.\right] = 2x^{(b_1+b_2)/2}\,K_{b_1-b_2}(2z^{1/2})$ \cite[Eq.~(6.5.8)]{Luke691} was used to derive \eqref{Eq.CaseI-2} with $K_{\alpha}(x)$ being the modified Bessel function of the second kind, and where $K_{-\frac{1}{2}}(x) = \sqrt{\frac{\pi}{2x}}\exp(-x)$ \cite[Eq.~(10.39.2)]{Olver10} was exploited to reach \eqref{Eq.CaseI-3}. This result is exactly the same as that reported in \cite[Eq.~(3.1)]{RathieJAP92}. Moreover, it is now evident that $\lim_{x \to 0}f_V(x) = \lim_{x \to 0}\exp(-x) = 1$, such that the constant $c$ involved in Corollary~\ref{Corollary_Limit} equals unity.

On the other hand, in the case of $d=1$, the general CDF given by \eqref{Eq-CDF} reduces to
\begin{eqnarray}
F_V(x)
&  =  &  \frac{1}{\sqrt{\pi}} \, \mathrm{G}_{1,\, 3}^{2,\, 1}\left[ \frac{x^2}{4} \left \vert \begin{gathered}  1 \\  \frac{1}{2}, \, 1, \, 0 \end{gathered}\right.\right]  \label{Eq.CaseI-4} \\
&  =  &  \left(\frac{\pi x}{2}\right)^{\frac{1}{2}}  \left[I_{\frac{1}{2}}(x) - \boldsymbol{L}_{\frac{1}{2}}(x)\right] \label{Eq.CaseI-5} \\
&  =  &  \sinh(x)-\cosh(x)+1   \label{Eq.CaseI-6} \\
&  =  & 1-\exp(-x),                 \label{Eq.CaseI-7}
\end{eqnarray}
where $\mathrm{G}_{1,\, 3}^{2,\, 1}\left[ x \left \vert ^{~~a+1/2} _{a, \, a+1/2, \, b} \right.\right] = \pi x^{(a+b)/2} [I_{a-b}(2x^{1/2}) - \boldsymbol{L}_{a-b}(2x^{1/2})]$ \cite[Eq.~(6.5.35)]{Luke691} was exploited to derive \eqref{Eq.CaseI-5}, with $I_{\alpha}(x)$ being the modified Bessel function of the first kind and $\boldsymbol{L}_{\alpha}(x)$ denoting the modified Struve function \cite[Eq.~(11.2.2)]{Olver10}. Then, $I_{\frac{1}{2}}(x) = \left(\frac{2}{\pi x}\right)^{1/2}\sinh(x)$ \cite[Eq.~(10.39.1)]{Olver10} and $\boldsymbol{L}_{\frac{1}{2}}(x) = \left(\frac{2}{\pi x}\right)^{1/2}(\cosh(x)-1)$ \cite[Eq.~(11.4.7)]{Olver10} were employed to reach \eqref{Eq.CaseI-6}. Finally, $\sinh(x) = \frac{1}{2}[\exp(x)-\exp(-x)]$ \cite[Eq.~(4.28.1)]{Olver10} and $\cosh(x) = \frac{1}{2}[\exp(x)+\exp(-x)]$ \cite[Eq.~(4.28.2)]{Olver10} were applied to get \eqref{Eq.CaseI-7}. Consequently, it is clear that \eqref{Eq.CaseI-7} is the CDF corresponding to the PDF given by \eqref{Eq.CaseI-3} .

\item{\textbf{Case II ($\boldsymbol{d=2}$)}:} In such a case, as per Table~\ref{Table_ParametersAB}, we get $A=3/\sqrt{\pi}$ and $B=4\pi^2/27$. Accordingly, the general PDF given by \eqref{Eq-PDF} simplifies to
\begin{eqnarray}
f_V(x)
&  =  &  \frac{3}{x\sqrt{\pi}} \, \mathrm{G}_{2,\, 4}^{4,\, 0}\left[ \frac{4\pi^2}{27}x^2 \left \vert \begin{gathered}  \frac{3}{2}, \, \frac{3}{2} \\  1, \, \frac{5}{6}, \, \frac{7}{6}, \, \frac{9}{6} \end{gathered}\right.\right]  \label{Eq.CaseII-1} \\
&  =  &  \frac{3}{x\sqrt{\pi}} \, \mathrm{G}_{1,\, 3}^{3,\, 0}\left[ \frac{4\pi^2}{27}x^2 \left \vert \begin{gathered}  \frac{3}{2} \\  1, \, \frac{5}{6}, \, \frac{7}{6} \end{gathered}\right.\right]  \label{Eq.CaseII-2} \\
&  =  &  \frac{8}{9}{\pi}x K_{\frac{1}{6}}^{2}\left(\frac{2\pi x}{3\sqrt{3}}\right),    \label{Eq.CaseII-3}
\end{eqnarray}
where the order reduction property of the Merjer's $\mathrm{G}$-function \cite[Eq.~(2.2.3)]{Mathai93} was employed to derive \eqref{Eq.CaseII-2} from \eqref{Eq.CaseII-1}, and $\mathrm{G}_{1,\, 3}^{3,\, 0}\left[ x \left \vert ^{~~a+1/2} _{a+b, \, a-b, \, a} \right.\right] = 2\pi^{-1/2} x^{a} K_{b}^{2}(x^{1/2}) $ \cite[Eq.~(6.5.36)]{Luke691} was exploited to reach \eqref{Eq.CaseII-3}, which is in accordance with that reported in \cite[Eq.~(3.2)]{RathieJAP92}.

On the other hand, in the case of $d=2$, the general CDF shown in \eqref{Eq-CDF} reduces to
\begin{eqnarray}
F_V(x)
\hspace{-5pt}&  =  & \hspace{-5pt}  \frac{3}{2\sqrt{\pi}} \, \mathrm{G}_{3,\, 5}^{4,\,1}\left[ \frac{4\pi^2}{27}x^2 \left \vert \begin{gathered}  1, \,  \frac{3}{2}, \, \frac{3}{2} \\  1, \, \frac{5}{6}, \, \frac{7}{6}, \, \frac{9}{6}, \, 0 \end{gathered}\right.\right]  \label{Eq.CaseII-4} \\
&  =  & \hspace{-5pt} \frac{3}{2\sqrt{\pi}} \, \mathrm{G}_{2,\, 4}^{3,\,1}\left[ \frac{4\pi^2}{27}x^2 \left \vert \begin{gathered}  1, \,  \frac{3}{2} \\  1, \, \frac{5}{6}, \, \frac{7}{6}, \, 0 \end{gathered}\right.\right],  \label{Eq.CaseII-5}
\end{eqnarray}
where the order reduction property of the Merjer's $\mathrm{G}$-function \cite[Eq.~(2.2.3)]{Mathai93} was used again to derive \eqref{Eq.CaseII-5} from \eqref{Eq.CaseII-4}.

\item{\textbf{Case III ($\boldsymbol{d=3}$)}:} Now, by virtue of Table~\ref{Table_ParametersAB}, we have $A=560\sqrt{2}/(81\pi)$ and $B=27\pi^2/16$. Accordingly, the general PDF given by \eqref{Eq-PDF} reduces to

\vspace{-10pt}
{\scriptsize
\begin{eqnarray}
f_V(x)
\hspace{-5pt}&  =  &\hspace{-5pt}  \frac{560\sqrt{2}}{81\pi x} \, \mathrm{G}_{4, \, 6}^{6, \, 0}\left[ \frac{27\pi^2}{16}x^2 \left \vert \begin{gathered}  \frac{11}{6}, \, \frac{13}{6}, \, 2, \, 2
\\  1, \, \frac{3}{2}, \, \frac{10}{8}, \, \frac{12}{8}, \, \frac{14}{8}, \, \frac{16}{8} \end{gathered}\right.\right]  \label{Eq.CaseIII-1} \\
&  =  & \hspace{-5pt} \frac{560\sqrt{2}}{81\pi x} \, \mathrm{G}_{3, \, 5}^{5, \, 0}\left[ \frac{27\pi^2}{16}x^2 \left \vert \begin{gathered}  \frac{11}{6}, \, \frac{13}{6}, \, 2
\\  1, \, \frac{3}{2}, \, \frac{10}{8}, \, \frac{12}{8}, \, \frac{14}{8} \end{gathered}\right.\right],  \label{Eq.CaseIII-2}
\end{eqnarray}
}
\hspace{-5pt}where the order reduction property of Merjer's $\mathrm{G}$-function \cite[Eq.~(2.2.3)]{Mathai93} was used again to derive \eqref{Eq.CaseIII-2} from \eqref{Eq.CaseIII-1}.

\begin{figure}[t]
\centering
\includegraphics [width=3.75in, clip, keepaspectratio]{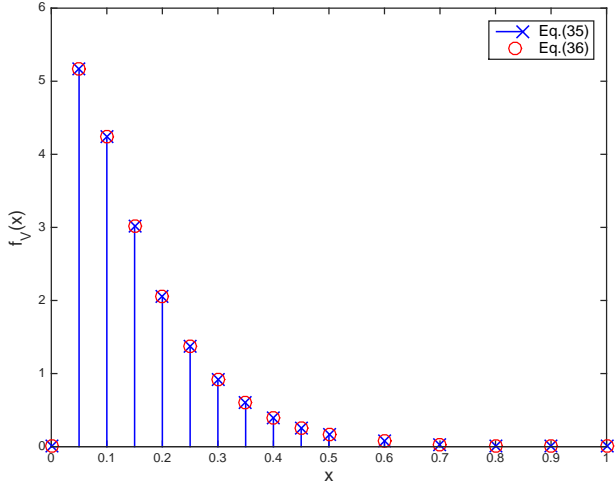}
\caption{Numerical comparison of the PDFs given by the analytical formula \eqref{Eq.CaseIII-2} obtained in this paper and by the integral expression \eqref{Eq.PDF-Muche} reported in \cite{MucheJSP96}.}
\label{Fig_PDFd3}
\end{figure}

In the case of $d=3$, an extremely complicated PDF expression of the volume of a typical cell was derived in \cite[Eq. (3.4)]{RathieJAP92}, which involves three infinite series consisting of Gamma, Psi and Zeta functions. Alternatively, a three-fold integral expression was reported in \cite{MucheJSP96}, which, for comparison purposes, is reproduced in Eq.~\eqref{Eq.PDF-Muche} on the top of the next page, where $g(\theta_1, \theta_2) = 1/\left(\sin{\frac{\theta_1}{2}} \sin{\frac{\theta_2}{2}} \sin{\frac{\theta_1 + \theta_2}{2}} \right)$.

\newcounter{mytempeqncnt3}
\begin{figure*}[!t]
\setcounter{mytempeqncnt3}{\value{equation}}
\setcounter{equation}{35}
\begin{equation} \label{Eq.PDF-Muche}
f_V(x)
= \frac{35x}{2}\int_{0}^{2\pi}\int_{0}^{2\pi-\theta_1}\int_{0}^{\pi}
{\sin{\theta_3} \, \exp\left(\frac{-2{\pi}x \, g(\theta_1, \theta_2)}{(1+\cos{\theta_3})\sin^{2}{\theta_3}}\right)} \, \mathrm{d}\theta_3 \,\mathrm{d}\theta_2 \,\mathrm{d}\theta_1.
\end{equation}
\setcounter{equation}{\value{mytempeqncnt3}}
\end{figure*}
\setcounter{equation}{36}

Figure~\ref{Fig_PDFd3} compares the numerical results of \eqref{Eq.CaseIII-2} and \eqref{Eq.PDF-Muche}. It is seen that they agree perfectly with each other. However, in comparison with the complicated triple integration of \eqref{Eq.PDF-Muche} which converges very slowly, computing \eqref{Eq.CaseIII-2} is very fast by using the built-in Meijer's $\mathrm{G}$-funcion in popular numerical software like Mathematica$^\circledR$. More importantly, the compact closed-form expression \eqref{Eq.CaseIII-2} enables further mathematically tractable processing. Finally, it is remarkable that, as previously pointed out in Corollary~\ref{Corollary_Limit}, since \eqref{Eq.CaseIII-2} is singular at the particular point $x = 0$, the value of $x=10^{-6}$ was used to approach this singularity in the numerical experiments. At the latter point, the numerical results of \eqref{Eq.CaseIII-2} and \eqref{Eq.PDF-Muche} are identical, being $6.84884 \times 10^{-4}$.

\item{\textbf{Case IV ($\boldsymbol{d \ge 4}$)}:}  In the open literature, no probability distribution of the volume of a typical cell of Poisson-Delaunay triangulations was ever reported for the case of $d \ge 4$. In light of Theorem~\ref{Theorem1}, however, we can easily obtain the probability distribution in any case of $d \ge 4$. For example, the PDFs of the volume in the case of $d=4$ and $d=5$ can be explicitly given by

\vspace{-10pt}
{\footnotesize
\begin{equation}  \label{Eq.CaseIV-1}
f_V(x)
=  \frac{A}{x} \, \mathrm{G}_{5, \, 7}^{7, \, 0}\left[ Bx^2 \left \vert \begin{gathered}  \frac{9}{4}, \, \frac{10}{4}, \, \frac{11}{4}, \, \frac{5}{2}, \, \frac{5}{2}
\\  1, \, \frac{3}{2}, \, 2, \, \frac{17}{10}, \, \frac{19}{10}, \, \frac{21}{10}, \, \frac{23}{10} \end{gathered}\right.\right],
\end{equation}
}
and
{\scriptsize
\begin{equation}  \label{Eq.CaseIV-2}
f_V(x)
=  \frac{A}{x} \, \mathrm{G}_{7, \, 9}^{9, \, 0}\left[ Bx^2 \left \vert \begin{gathered}  \frac{26}{10}, \, \frac{27}{10}, \, \frac{28}{10}, \, \frac{29}{10}, \, 3, \, 3,\, 3
\\  1, \, \frac{3}{2}, \, 2,\, \frac{5}{2}, \, \frac{26}{12}, \, \frac{28}{12}, \, \frac{30}{12}, \, \frac{32}{12}, \, \frac{34}{12} \end{gathered}\right.\right],
\end{equation}
}
\hspace{-1ex}respectively, where the values of constants $A$ and $B$ pertaining to each case are available from Table~\ref{Table_ParametersAB}.
\end{enumerate}

\begin{figure}[t]
	\centering
	\includegraphics [width=3.75in, clip, keepaspectratio]{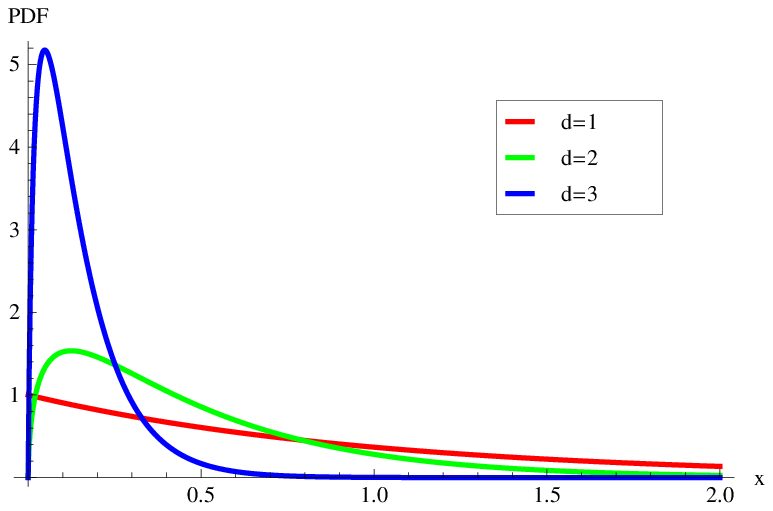}
	\caption{PDF of the volume of a typical Poisson-Delaunay cell in Euclidean space $\mathbb{R}^{d}$, for $d =1, 2, 3$.}
	\label{Fig_PDFd123}
\end{figure}

\begin{figure}[t]
	\centering
	\begin{subfigure}[b]{.5\textwidth}
		\includegraphics[width=.95\textwidth]{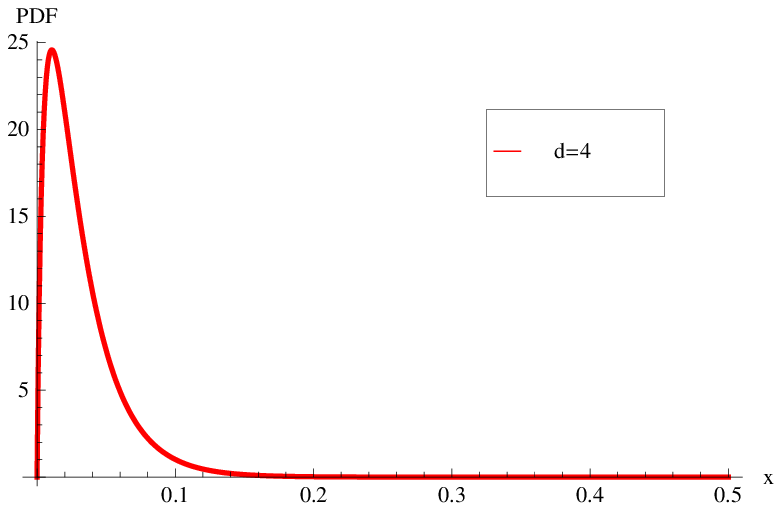}
		\caption{The case with $d=4$.}
	\end{subfigure}
	%add desired spacing between images, e. g. ~, \quad, \qquad etc, or a blank line to force the subfigure onto a new line.
	\begin{subfigure}[b]{.5\textwidth}
		\includegraphics[width=.95\textwidth]{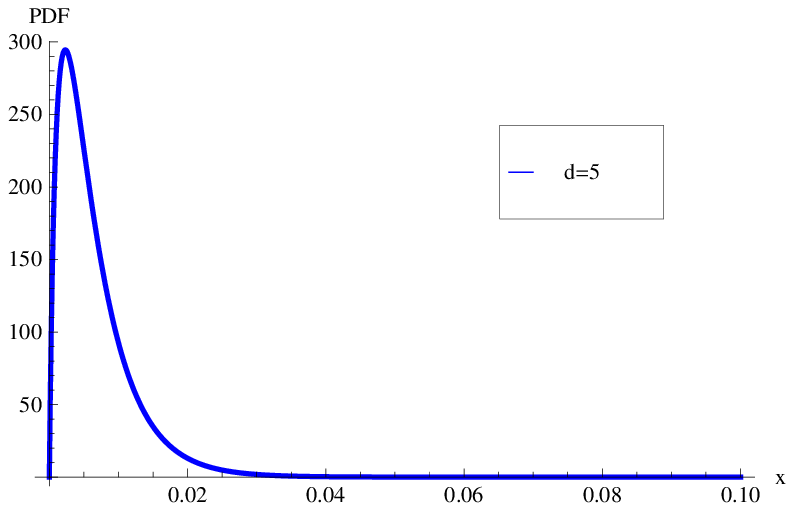}
		\caption{The case with $d=5$.}
	\end{subfigure}
	\caption{PDF of the volume of a typical Poisson-Delaunay cell in Euclidean space $\mathbb{R}^{d}$, for $d =4, 5$.}
	\label{Fig_PDFd45}
\end{figure}

\begin{table*}[t]
\caption{Shape characteristics of the PDF of the volume of a typical Poisson-Delaunay cell.}
\begin{center}
\begin{tabular}{cccccc}
\toprule
$d$            & 1                 & 2                   & 3                        & 4                              & 5     \\
\midrule\midrule
Mean         & $1$             &$0.5$             &$0.14776$          &$0.0314685$            &$0.0053551$        \\
\midrule
Variance    & $1$             &$0.19328$     &$0.0153653$       &$0.000694902$       &$0.00002089$       \\
\midrule
Skewness  & $2$             &$1.82424$     &$1.8045$            &$1.86166$               &$1.95854$            \\
\midrule
Kurtosis     & $6$             &$5.05614$     &$5.03457$           &$5.46232$               &$6.16637$             \\
\bottomrule
\end{tabular}
\end{center}
\label{Table_ShapeCharacteristics}
\end{table*}

\subsection{The Shape Characteristics of the PDF}
Now, the PDF expressions obtained in \eqref{Eq.CaseI-1}, \eqref{Eq.CaseII-1}, \eqref{Eq.CaseIII-1}, \eqref{Eq.CaseIV-1} and \eqref{Eq.CaseIV-2} are applied to numerically generate the distribution functions of the volume of a typical cell of Poisson-Delaunay triangulations in $\mathbb{R}^{d}$, with $d = 1, 2, 3, 4, 5$, respectively. For better clarity of illustration, the PDFs of the volume in the cases of $d = 1, 2, 3$ are collectively shown in Fig.~\ref{Fig_PDFd123} whereas the PDFs of the volume in the case of $d=4, 5$ are separately depicted in two subfigures of Fig.~\ref{Fig_PDFd45}. Notice that the scale of the horizontal axis of Figs.~\ref{Fig_PDFd123} and~\ref{Fig_PDFd45} becomes smaller and smaller whereas the scale of the vertical axis turns larger and larger, although the shape of curves looks similar to each other. On the other hand, the mean, variance, skewness and kurtosis of the PDF pertaining to each case under consideration are computed and summarized in Table~\ref{Table_ShapeCharacteristics}.

It is observed from Figs.~\ref{Fig_PDFd123} and \ref{Fig_PDFd45} that all PDF curves are unimodal, highly skewed right, and leptokurtic. In particular, in the case of $d=1$, the mode is at the leftmost endpoint $x=0$.  In the case of $d>1$, the mode is positive but shifts from right to left as the value of $d$ increases. It is noteworthy that, given the moments computed by \eqref{Eq-moments}, the unimodality of the corresponding PDF can be readily proven by using the sufficient and necessary condition shown in \cite[Theorem~3]{Johnson51}. These visible shape characteristics of the PDF of the volume are well supported by the statistics summarized in Table~\ref{Table_ShapeCharacteristics}. More specifically, the value of skewness being larger than unity means highly right skewness, and the value of kurtosis being larger than $3$ implies leptokurtic distribution.

\section{Application}
\label{Apps}

Delaunay triangulations find many applications in various scientific fields. Here, we take three of its  properties and the corresponding applications as example. In particular, in the planar case, i.e., $d=2$, the Delaunay triangulation has a striking advantage: among all possible triangulations of a point set, the Delaunay triangulation maximizes the minimum angle.  In planar or any higher dimensional cases and over all possible triangulations of a point set, the Delaunay triangulation minimizes the maximum enclosing radius of any simplex, where the enclosing radius of a simplex is defined as the minimum radius of an enclosing sphere \cite{RajanDCG9402}. These properties are widely applied for mesh generation in computer graphics \cite{Cheng13}. A third property of the Delaunay triangulation is ray-shoot monotone, which is very useful in the area of visibility and ordering multivariate data  \cite[Section~6.6]{Okabe00}. Using the analytical distribution functions of the volume of a typical cell shown in Theorem~\ref{Theorem1}, more statistics on these applications become mathematically tractable even in spatial or higher dimensional Euclidean space, which were usually obtained by simulation experiments. For more information on various applications of Delaunay triangulations, the interested reader is referred to \cite{Okabe00} and \cite{Chiu13}.

\begin{figure}[t]
	\centering
	\includegraphics [width=3.75in, clip]{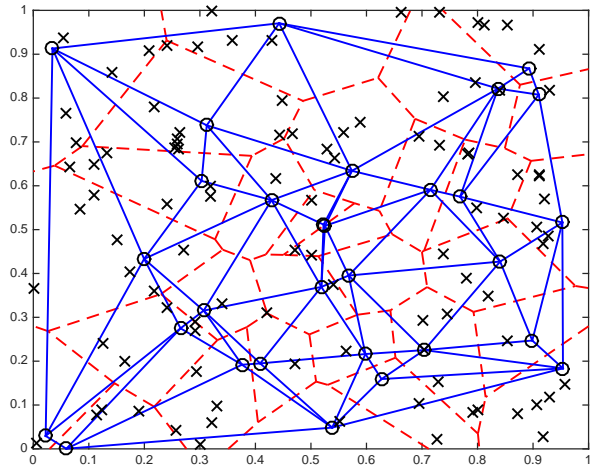}
	\caption{An illustrative cellular network modeled by the Poisson-Voronoi tessellation (the polygons with red dash boundaries) or by the dual Poisson-Delaunay triangulation (the triangles with blue solid boundaries), where the black circles (`$\circ$') refer to the BSs and the cross marks (`$\times$') denote the UEs, with normalized coverage area of one squared kilometers.}
	\label{Fig_Tessellations}
\end{figure}

In the following, an application of Poisson-Delaunay triangulation to CoMP transmission in 5G wireless communications is elaborated, and the pertinent void cell effect is precisely evaluated by using the obtained distribution functions.

\subsection{A Novel CoMP Transmission Scheme Based on Poisson-Delaunay Triangulation}
In conventional cellular networks, each user equipment (UE) is associated with its nearest base station (BS) and gets service from this BS. Accordingly, the coverage area of a network is essentially divided into adjacent polygonal cells, as shown with red dash boundaries in Fig.~\ref{Fig_Tessellations} where the BSs and UEs are denoted by the `$\circ$' and `$\times$' marks, respectively. In theory, this classic network structure can be effectively modelled by means of Poisson-Voronoi tessellation in the field of stochastic geometry \cite{Haenggi12}. To further improve the quality of service (QoS) of wireless networks, CoMP transmission is universally recognized as a promising technique for 5G wireless systems by academic researchers and standard bodies as well \cite{3GPPTR36.819, NigamTCOM1411, NguyenTWC1701}. For instance,  CoMP transmission was widely applied to 5G radio access design, like for dense networks in  \cite{ChenNETWORK1705} or for could-RANs (radio access networks) in \cite{RajannaWCL1702, HaTVT1609, Ferdouse17}. By using CoMP technique, multiple BSs are joined together to simultaneously serve a particular UE. To this end, a critical step to implement CoMP technique in practice is to determine which BSs should be coordinated for a specific UE. More formally, a cooperation set of BSs must be constructed prior to data transmission, e.g., by choosing the nearest one or more than one BSs in the sense of Euclidean distance. The construction of cooperation set is generally CSI dependent and time intensive.

Unlike the CoMP strategies in conventional Poisson-Voronoi cells where a {\it dynamic} mechanism has to be relied upon to determine and update the cooperation set of a UE, in the following we propose a novel CoMP mechanism based on Poisson-Delaunay triangulation, whereby the cooperation set of a UE is {\it fixed and uniquely determined} by the geometric locations of its nearby BSs. In principle, the Poisson-Delaunay triangulation is the dual graph for a Poisson-Voronoi tessellation. For instance, as illustrated in Fig.~\ref{Fig_Tessellations}, the Poisson-Delaunay triangulation dual to the Poisson-Voronoi tessellation shown in red dash boundaries consists of the triangles with blue solid boundaries. In practice, with the geometric locations of BSs, the Poisson-Delaunay triangulation dual to a Poisson-Voronoi tessellation is uniquely determined and can be efficiently constructed by using, e.g., the radial sweep algorithm or divide-and-conquer algorithm \cite[ch. 4]{Hjelle06}. Subsequently, for each UE, the CoMP cooperation set is determined as follows.

\begin{figure}[t]
	\centering
	\includegraphics [width=3.5in, clip, keepaspectratio]{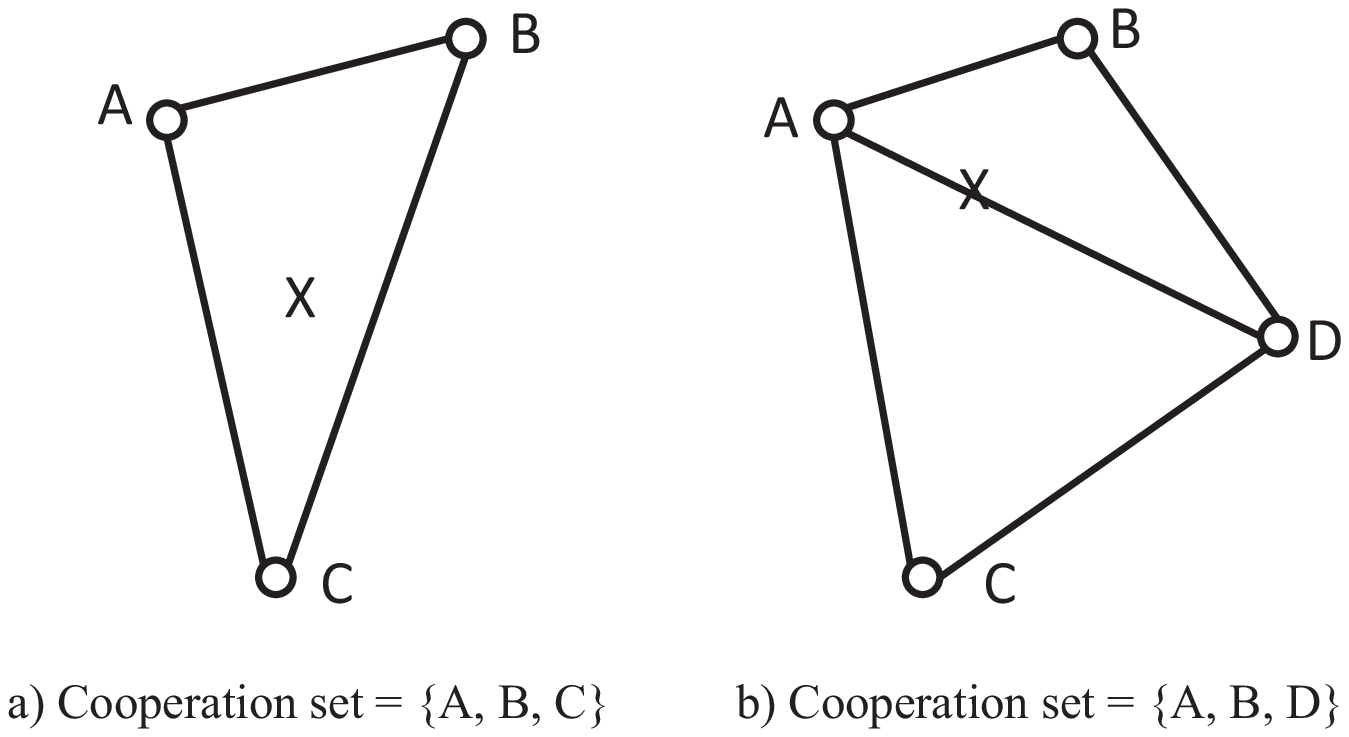}
	\caption{Principles to determine the CoMP cooperation set at each UE with respect to Poisson-Delaunay triangulation, where the black circles (`$\circ$') and cross marks (`$\times$') refer to BSs and UEs, respectively.}
	\label{Fig-2}
\end{figure}

As shown in Fig.~\ref{Fig-2}-a), if a UE is located inside a Poisson-Delaunay triangular cell, the three BSs at the vertices of the triangle are chosen and form the CoMP cooperation set. On the other hand, if a UE is exactly located on the edge of a triangle shown in Fig.~\ref{Fig-2}-b), there must be an adjacent triangle which shares the same edge and they both form a quadrilateral (incidentally, the edge effect of the whole cellular network is ignored due to its large coverage area). Among the four BSs at the vertices of the quadrilateral, the UE on the edge chooses the two BSs at both ends of the edge and a third BS which is closer to the UE between the remaining two opposite BSs, so as to form the CoMP cooperation set.

\begin{figure}[t]
	\centering
	\includegraphics [width=3.75in, clip, keepaspectratio]{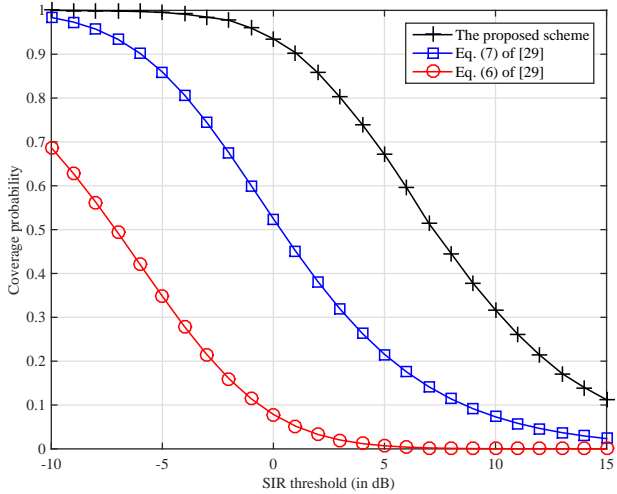}
	\caption{Coverage probability versus signal-to-interference ratio (SIR) threshold in the unit dB.}
	\label{Fig-CoverageProb}
\end{figure}

Compared with dynamic cooperation set, the proposed CoMP scheme based on Poisson-Delaunay triangulation has two main features: {\it i)} the cooperation BS set pertaining to a particular UE is fixed and can be offline determined once the geometric locations of BSs are known, and {\it ii)} the average coverage area of Poisson-Delaunay triangular cells is only half of that of dual Poisson-Voronoi polygonal cells, i.e., $1/{(2\lambda)}$ versus $1/{\lambda}$ with $\lambda$ being the intensity of BSs \cite[Tables 5.5.1 and 5.11.1]{Okabe00}. Clearly, the former feature facilitates the implementation of CoMP technique yet at the cost of coordination gain. The latter feature, on the other hand, benefits improving the coverage probability and spectral efficiency of wireless networks, like the technique known as small cells \cite{AndrewsJSAC1406}. For illustration purposes, Monte-Carlo simulation experiments are performed to evaluate the coverage probability of the proposed CoMP scheme. In particular, in line with \cite{JungCL1308}, the worst-case UEs at the vertices of conventional Poisson-Voronoi tessellation (cf. Fig.~1 of \cite{JungCL1308}) is chosen as the typical UE. As shown in Fig.~\ref{Fig-CoverageProb}, the coverage probability of our proposed scheme outperforms that with optimal point selection \cite[Eq.~(7)]{JungCL1308} or that without CoMP \cite[Eq.~(6)]{JungCL1308}.

In other words, compared with conventional CoMP strategies based on Poisson-Voronoi tessellation, the proposed scheme based on Poisson-Delaunay triangulation is more suitable for small cell networks. However, as the cell size becomes smaller and smaller, the effect of void cell emerges, as discussed below.

\subsection{The Void Cell Effect}
The void cell effect originates from the user-centric cell association, e.g., each UE in a cellular network associates with its nearest BS. As a toy example, we may consider that there are $m \times n$ UEs uniformly distributed in a unit-area network which is divided into $n$ equal-sized cells covered by BSs. For a particular cell, it is apparent that the probability that this cell contains no any UE is given by $\left(1-\frac{1}{n}\right)^{mn}$. Moreover, as the cell size becomes smaller or, equivalently, as the value of $n$ becomes larger while $m$ remains constant, we have $\lim_{n \to \infty} {\left(1-\frac{1}{n}\right)^{mn}} = e^{-m}$. Clearly, when the value of $m$, i.e., the ratio of the number of UEs to the number of BSs is small, the probability of void cell, i.e., $e^{-m}$, is not negligible. For instance, the probability of void cell is $e^{-3} \approx 5\%$ in the case of $m = 3$.

Without accounting for the void cell effect, the analyses of network performance metrics, like spectral efficiency, energy efficiency, coverage probability and network throughput, are inevitably underestimated since those BSs corresponding to void cells do not introduce any interference to adjacent cells. Despite the extreme importance of the void cell effect in small cell networks or, equivalently, dense networks, there are very few works in the literature that touch this issue and almost all prior works based on the modelling of stochastic geometry for wireless networks overlook this effect (see e.g., \cite{AndrewsTCOM1111, HuangTIT1304, LeeTIT1603, BjornsonTIT1603}), since exact evaluation is mathematically intractable. In particular, since the distribution functions of a typical cell of Poisson-Voronoi tessellation is still an open problem \cite[Section 9.7]{Chiu13}, exact evaluation of the void cell effect pertaining to Poisson-Voronoi cells is not feasible so far \cite{LiuICC15, LiuJSAC16}.

By using the previously obtained mathematical results, the void cell effect of the proposed CoMP based on Poisson-Delaunay triangulation, can be accurately evaluated as follows. Furthermore, a green power control strategy to exploit the void cell effect is developed.

By definition, the exact void-cell probability (say, $p$) of the CoMP scheme based on Poisson-Delaunay triangulation can be computed as
\begin{equation}  \label{Eq-VoidProbability-Triangle-1}
p = \mathcal{E}_{V} \left\{e^{-\lambda_{\rm UE} {V}}\right\} = \int_0^\infty{e^{-\lambda_{\rm UE} x} f_{V}(x)} \,{\rm d}x,
\end{equation}
where $\lambda_{\rm UE}$ denotes the intensity of UEs and $V$ represents the volume of a typical cell in the Poisson-Delaunay triangulation created by BSs. By virtue of the mathematical results obtained in this paper, including Theorem~\ref{Theorem1}, Corollary~\ref{Corollary_Intensity} and \eqref{Eq.CaseII-2}, Eq.~\eqref{Eq-VoidProbability-Triangle-1} can be explicitly calculated such that
\begin{align}
p
&   =  \int_0^\infty{\frac{3 e^{-\lambda_{\rm UE} x} }{x\sqrt{\pi}} \, \mathrm{G}_{1,\, 3}^{3,\, 0}\left[ \frac{4\pi^2}{27}\left(\lambda_{\rm BS} x\right)^2 \left \vert \begin{gathered}  \frac{3}{2} \\  1, \, \frac{5}{6}, \, \frac{7}{6} \end{gathered}\right.\right]} \,{\rm d}x   \label{Eq-VoidProbability-Triangle-2}  \\
&  =   \frac{3}{2 \pi} \, \mathrm{G}_{3,\, 3}^{3,\, 2}\left[ \frac{16 \pi^2 \lambda_{\rm BS}^2}{27 \lambda_{\rm UE}^2} \left \vert \begin{gathered}  \frac{1}{2}, \, 1, \, \frac{3}{2} \\  1, \, \frac{5}{6}, \, \frac{7}{6} \end{gathered}\right.\right],  \label{Eq-VoidProbability-Triangle-3}
\end{align}
where \cite[Vol. 3, Eq. (2.24.3.1)]{Prudnikov86} was exploited to derive from \eqref{Eq-VoidProbability-Triangle-2} to \eqref{Eq-VoidProbability-Triangle-3}, with $\lambda_{\rm BS}$ referring to the intensity of BSs.

By recalling the Jensen's inequality, on the other hand, we may obtain a lower bound on the void probability pertinent to the CoMP transmission. That is,
\begin{equation}  \label{Eq-VoidProbability-4}
p = \mathcal{E}_{V}\left\{e^{-\lambda_{\rm UE} {V}}\right\}
			\ge e^{-\lambda_{\rm UE} \, \mathcal{E}_ {V}\{ {V}\}}
			= \exp\left(-\frac{\lambda_{\rm UE}}{2 \, \lambda_{\rm BS}}\right),
\end{equation}
where the fact that the average coverage area of a typical triangular cell is $\mathcal{E}_{V}\{V\} = 1/(2\lambda_{\rm BS})$ \cite[Table 5.11.1]{Okabe00}, was employed to reach the last equality in \eqref{Eq-VoidProbability-4}.

\begin{figure}[t]
	\centering
	\includegraphics [width=3.75in, clip, keepaspectratio]{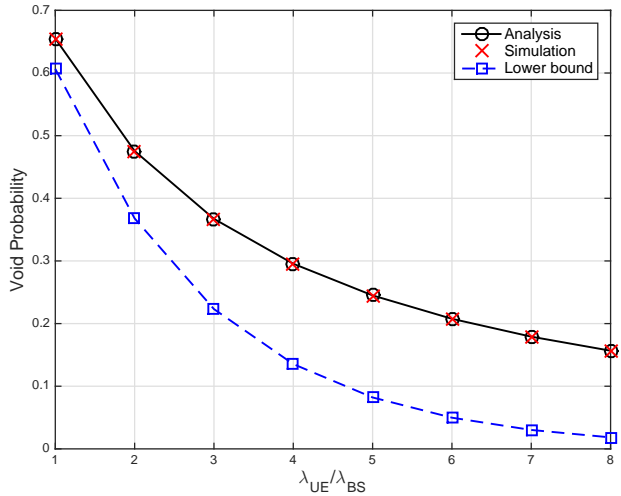}
	\caption{The void probability pertaining to the coordinated multi-point transmission based on Poisson-Delaunay triangulation.}
	\label{Fig_VoidCoMP}
\end{figure}

To demonstrate the effectiveness of preceding analysis, Fig.~\ref{Fig_VoidCoMP} illustrates the void probability versus the intensity ratio of UEs to BSs, i.e., $\lambda_{\rm UE}/\lambda_{\rm BS}$.  In the pertaining simulation experiments, the value of $\lambda_{\rm UE}/\lambda_{\rm BS}$ ranges from 1 to 8, i.e., the network become denser and denser. As seen from the figure, the void probability is always greater than $0.1$ in the region of $\lambda_{\rm UE}/\lambda_{\rm BS}$ of interest, which is evidently not negligible in practice. For instance, when $\lambda_{\rm UE}/\lambda_{\rm BS} = 5$, the void probability of is as large as $0.25$. On the other hand, the void cell probability decreases with larger $\lambda_{\rm UE}/\lambda_{\rm BS}$, as expected. Moreover, Fig.~\ref{Fig_VoidCoMP} illustrates that the simulation results are in full agreement with the numerical results computed according to \eqref{Eq-VoidProbability-Triangle-3}, whereas the lower bounds computed as per \eqref{Eq-VoidProbability-4} are not tight, owing to the impreciseness of the Jensen's inequality.

To deal with such a high void probability caused by smaller cell size, the so-called green power control can be applied at BSs, whereby a BS may switch between `active mode' with high power consumption ($P_{\rm A}$) and `sleeping mode' with very low power consumption ($P_{\rm S}$). In particular, the BSs corresponding to void cells may operate according to the `sleeping mode'.\footnote{It is noteworthy that the `sleeping mode' does not necessarily imply that a whole BS goes into sleep, but that only the part of Tx antennas aiming at a particular cell switches to `sleeping mode'. For instance, a large-scale antenna array at a BS can form multiple directional beams and serve different cells independently.} This power-control strategy will yield at least two major advantages. One is lower inter-cell interference, resulting in better network performance. The other is lower average power consumption (denoted $\bar{P})$ at a BS, which can be characterized by
\begin{equation}  \label{Eq-GPC}
\bar{P} = (1-p) \, P_{\rm A} + p \, P_{\rm S} \approx (1-p) \, P_{\rm A},
\end{equation}
where $p$ denotes the void probability given by \eqref{Eq-VoidProbability-Triangle-3}. Eq.~\eqref{Eq-GPC} implies that the average power consumption at a BS is proportional to the complementary void probability.

\section{Concluding Remarks}
\label{Section:Conclusions}
As a leading feature of a typical cell of Poisson-Delaunay triangulations, the probability distributions of its volume were developed in an analytical and unified way, and are applicable to Euclidean spaces of arbitrary dimension. The main results are compact and elegant, and can be efficiently computed in a numerical way by using built-in functions of popular numerical softwares. In particular, the shape characteristics of the probability density function of the volume were demonstrated to be unimodal, highly skewed right, and leptokurtic. The results obtained shed new light on the development of unified theory on the characteristics of a typical cell, and enable further mathematically tractable processing for the application of Poisson-Delaunay triangulations to various scientific disciplines, such as coordinated multi-point transmission in wireless networks.

\bibliographystyle{IEEEtran}
\bibliography{References}

\vfill 

\begin{IEEEbiography}
 [{\includegraphics[width=1in, height=1.25in, clip, keepaspectratio]{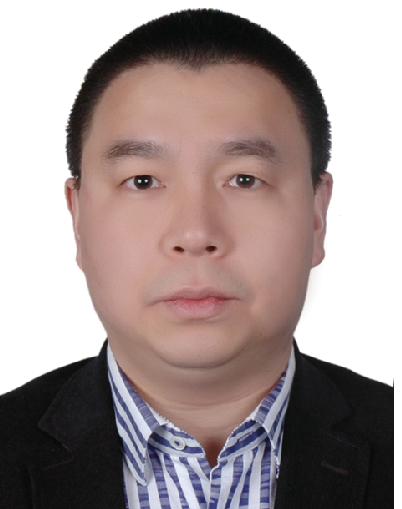}}]{Minghua Xia} (M'12) obtained his Ph.D. degree in Telecommunications and Information Systems from Sun Yat-sen University, Guangzhou, China, in 2007. Since 2015, he has been a Professor with Sun Yat-sen University.

From 2007 to 2009, he was with the Electronics and Telecommunications Research Institute (ETRI) of South Korea, Beijing R\&D Center, Beijing, China, where he worked as a member and then as a senior member of engineering staff and participated in the projects on the physical layer design of 3GPP LTE mobile communications. From 2010 to 2014, he was in sequence with The University of Hong Kong, Hong Kong, China; King Abdullah University of Science and Technology, Jeddah, Saudi Arabia; and the Institut National de la Recherche Scientifique (INRS), University of Quebec, Montreal, Canada, as a Postdoctoral Fellow. His research interests are in the general area of 5G wireless communications, and in particular the design and performance analysis of multi-antenna systems, cooperative relaying systems and cognitive relaying networks, and recently focus on the design and analysis of wireless power transfer and/or energy harvesting systems, as well as massive MIMO and small cells. He holds two patents granted in China.

 Dr. Xia received the Professional Award at the IEEE TENCON, held in Macau, in 2015. He was also recognized as an Exemplary Reviewer by {\scshape IEEE Transactions on Communications} in 2014, {\scshape IEEE Communications Letters} in 2014, and {\scshape IEEE Wireless Communications Letters} in 2014 and 2015.
 \end{IEEEbiography}

\begin{IEEEbiography}
[{\includegraphics[width=1in,height=1.25in,clip,keepaspectratio]{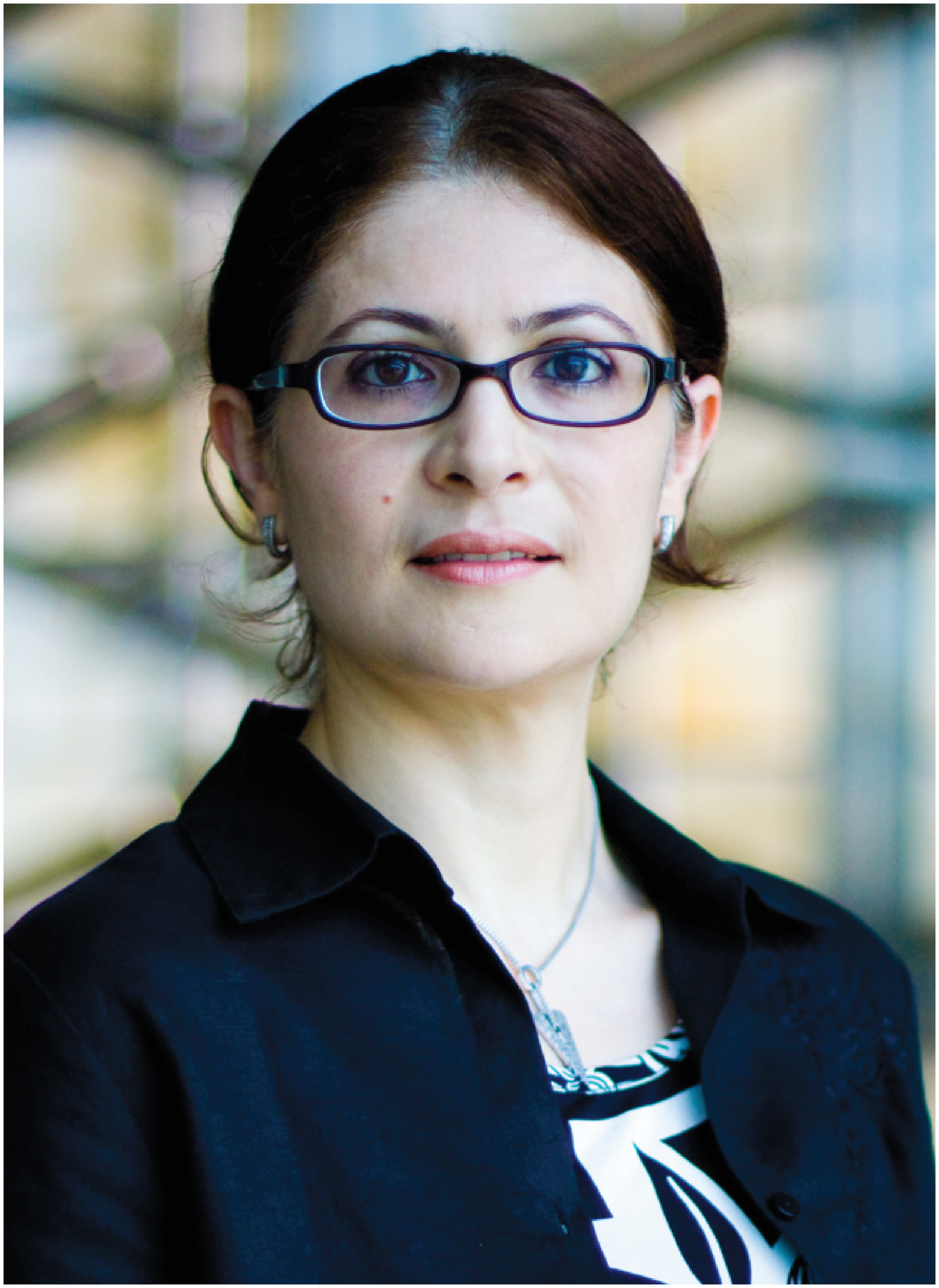}}]
{Sonia A\"{\i}ssa} (S'93-M'00-SM'03) received her Ph.D. degree in Electrical and Computer Engineering from McGill University, Montreal, QC, Canada, in 1998. Since then, she has been with the Institut National de la Recherche Scientifique-{\it Energy, Materials and Telecommunications} Center (INRS-EMT), University of Quebec, Montreal, QC, Canada, where she is a Full Professor.

From 1996 to 1997, she was a Researcher with the Department of Electronics and Communications of Kyoto University, and with the Wireless Systems Laboratories of NTT, Japan. From 1998 to 2000, she was a Research Associate at INRS-EMT. In 2000-2002, while she was an Assistant Professor, she was a Principal Investigator in the major program of personal and mobile communications of the Canadian Institute for
Telecommunications Research, leading research in radio resource management for wireless networks. From 2004 to 2007, she was an Adjunct Professor with Concordia University, Montreal. She was Visiting Invited Professor at Kyoto University, Japan, in 2006, and Universiti Sains Malaysia, in 2015. Her research interests include the modeling, design and performance analysis of wireless communication systems and networks.

Dr. A\"{\i}ssa is the Founding Chair of the IEEE Women in Engineering Affinity Group in Montreal, 2004-2007; acted as TPC Symposium Chair or Cochair at IEEE ICC '06 '09 '11 '12; Program Cochair at IEEE WCNC 2007; TPC Cochair of IEEE VTC-spring 2013; and TPC Symposia Chair of IEEE Globecom 2014. Her main editorial activities include: Editor, {\scshape IEEE Transactions on Wireless Communications}, 2004-2012; Associate Editor and Technical Editor, {\scshape IEEE Communications Magazine}, 2004-2015; Technical Editor, {\scshape IEEE Wireless Communications Magazine}, 2006-2010; and Associate Editor, {\it Wiley Security and Communication Networks Journal}, 2007-2012. She currently serves as Area Editor for the {\scshape IEEE Transactions on Wireless Communications}. Awards to her credit include the NSERC University Faculty Award in 1999; the Quebec Government FRQNT Strategic Faculty Fellowship in 2001-2006; the INRS-EMT Performance Award multiple times since 2004, for outstanding achievements in research, teaching and service; and the Technical Community Service Award from the FQRNT Centre for Advanced Systems and Technologies in Communications, 2007. She is co-recipient of five IEEE Best Paper Awards and of the 2012 IEICE Best Paper Award; and recipient of NSERC Discovery Accelerator Supplement Award. She is a Distinguished Lecturer of the IEEE Communications Society (ComSoc) and an Elected Member of the ComSoc Board of Governors. Professor A\"{\i}ssa is a Fellow of the Canadian Academy of Engineering.
\end{IEEEbiography}

\end{document}